\documentclass[a4paper,aps,prc,twocolumn,amsmath,amssymb,showkeys]{revtex4}
\usepackage{graphicx}
\usepackage{mathptmx}
\usepackage{epsfig}
\usepackage{units}
\usepackage{url}
\usepackage{upgreek}
\usepackage{wasysym}
\usepackage{hyperref}

\hypersetup{%
  pdfstartview=FitH,%
  bookmarksopen=true,%
  breaklinks=true,%
  colorlinks=true,%
  linkcolor=blue,anchorcolor=blue,%
  citecolor=blue,filecolor=blue,%
  menucolor=blue,%
  urlcolor=blue%
}

\begin{document}

\title{The SiRi Particle-Telescope System}

\author{M.~Guttormsen}
\email[Email address: ]{magne.guttormsen@fys.uio.no}
\affiliation{Department of Physics, University of Oslo, Norway}
\author{A.~B\"urger}
\affiliation{Department of Physics, University of Oslo, Norway}
\author{T.E.~Hansen}
\affiliation{SINTEF, Department for Microsystems and Nanotechnology, Oslo, Norway}
\author{N.~Lietaer}
\affiliation{SINTEF, Department for Microsystems and Nanotechnology, Oslo, Norway}

\begin{abstract}
  A silicon particle-telescope system for light-ion
  nuclear reactions is described.
  In particular, the system is designed to be optimized for level
  density and $\gamma$-ray strength function measurements with the
  so-called Oslo method.
  Eight trapezoidal modules are mounted at $\unit[5]{cm}$ distance
  from the target, covering 8~forward angles between $\theta =
  40$ and $54^{\circ}$.
  The thin front $\Delta E$ detectors ($\unit[130]{\upmu m}$) are
  segmented into eight pads, determining the reaction angle $\theta$
  for the outgoing charged ejectile.
  Guard rings on the thick back $E$ detectors ($\unit[1550]{\upmu m}$)
  guarantee low leakage current at high depletion voltage.
\end{abstract}

\maketitle

\section{Introduction}

The experimental nuclear physics group at the Oslo Cyclotron
Laboratory (OCL) has, through the last decades, investigated the
excitation energy region between quantum-order and chaos in nuclei.
The group has developed the so-called Oslo method \cite{OsloMethod},
which gives the number of energy levels accessible for the nucleus, as
well as the $\gamma$-ray strength function from these energetic
quantum states.

The OCL group has gained international renown and much attention for
its discoveries, see e.g.~\cite{SnPaper,Sc45Paper} and references
therein.
The most important results are ({\em i}) experimental evidence for
breaking of Bardeen-Cooper-Schrieffer (BCS) pairs and the melt down of
pair correlations in the nucleus, ({\em ii}) measurements of nuclear
heat capacity, ({\em iii}) discovery of a scissors-like vibration mode
and determination of the nature of its electromagnetic decay, ({\em
  iv}) discovery of enhanced low-energetic $\gamma$-emission in light
nuclei, and ({\em v}) measurements of vibrations of the nucleus'
neutron skin.
These discoveries are essential for astrophysical applications, and in
particular for the understanding of the distribution of elements in
our solar system.
The results are also important for transmutation of radioactive waste,
such that its lifetime can be dramatically reduced.

The experimental studies are based on in-beam coincidences between
$\gamma$-rays and charged reaction ejectiles.
The set-up includes an array of 28 $5^"\!\times 5^"$ NaI $\gamma$-ray
detectors (CACTUS) with a total efficiency of \unit[15]{\%}, and a set
of silicon particle telescopes.
Using standard $\Delta E-E$ silicon detectors, only eight
particle-telescopes could be fitted around the target inside the
CACTUS target chamber because of space constraints.
Therefore, the active detector area and, consequently, the detection
efficiency were small, calling for a replacement by modern
user-designed detectors whose properties will be described in the
following.

We foresee that the new silicon ring (SiRi) will lead to more
discoveries as fine structures in the data such as spin dependencies
can be studied.
We give a short outline of the design requirements in
section~\ref{sec:design}, and in section~\ref{sec:layout} the silicon
chip processes are described.
The signal handling and acquisition system are discussed in
section~\ref{sec:electronics}.
Finally, test results and conclusions are presented in
sections~\ref{sec:performance} and \ref{sec:conclusion},
respectively.

\section{Design parameters}
\label{sec:design}

The goal of the new particle-telescope system is to obtain a compact
set-up with high particle-$\gamma$ coincidence efficiency.
Compared to the previous detector set-up with conventional silicon
detectors, the goal was to obtain ten times higher efficiency without
degrading the particle energy resolution or the timing properties.

The detector telescopes are designed for the measurement of energy,
time, and to discriminate between different charged ejectiles from
light transfer or scattering reactions.
Typically, such nuclear reactions are (p,p$^\prime$), (p,d) and
(\textsuperscript{3}He,$\upalpha$), but also two-nucleon transfer
reactions like (p,$\upalpha$)~\cite{Sc43Paper} and
(p,t)~\cite{Zr9092Paper}.
Beam energies used are between $15$ and $\unit[45]{MeV}$.
The Oslo method requires that the reaction includes exactly one
outgoing charged particle.
Our main interest is to measure the direct reaction product, usually
in forward direction.
Therefore, to avoid particle pile-up events within one and the same
detector, special attention has to be given to the high elastic cross
section at low scattering angles

The input basis for the Oslo method is a set of $\gamma$-ray spectra
for all excitation energy bins $E_x$ between the ground state up to
the neutron separation energy $S_n$.
However, in order to determine $E_x$ accurately enough ($\Delta E_x <
\unit[200]{keV}$), it is not sufficient to know the beam energy,
reaction $Q$-value, and the energy of the outgoing particle.
The recoil energy of the daughter nucleus also depends on the
scattering angle $\theta$ between beam axis and ejectile, and thus, is
directly connected to the determination of $E_x$.
The recoil correction is of particular importance for lighter nuclei
and makes it necessary to measure $\theta$ with an uncertainty of
typically less than $\unit[\pm 1]{^{\circ}}$.

Pile-up events and accurate excitation energy measurements require a
certain granularity of the detectors.
However, to avoid possible misalignments and bad overlap between the
respective $\Delta E$ and $E$ pads, and at the same time to keep the
costs at a reasonable level, only the $\Delta E$ detectors were
segmented.
By requiring that only one $\Delta E$ pad fires, pile-up events in the
$E$ detector shared by the pads can be rejected.

The particle-telescopes are to be placed inside the existing vacuum
target chamber of the CACTUS NaI array.
The 28 NaI detectors are placed at a distance of $\unit[22]{cm}$ from
the target and are distributed on a spherical frame.
Each NaI is equipped with a conical $\unit[10]{cm}$ thick lead
collimator between the target and detector with an opening of
$\diameter = \unit[70]{mm}$ at the NaI-detector front surface.
The chamber is a cylindrical tube with an inner length
of~\unit[48.0]{cm} and a diameter of~\unit[11.7]{cm}.
To obtain reasonable high direct reaction cross sections with low spin
transfer, we measure the outgoing particles at angles  $\theta
= 47^{\circ} \pm \unit[7]{^{\circ}}$ with respect to the beam axis.
Lower scattering angles would give significant pile-up due to the
strongly increasing elastic cross section and, thus, impose the
necessity to run with lower beam current.

The center of each detector module is placed at \unit[5.0]{cm} from
the target.
Present technology requires that the silicon wafers are flat, and we
find that eight trapezoidal-shaped telescope modules form an
approximate ring around target.
The $\Delta E$ detectors is segmented into eight curved pads, covering
mean scattering angles $\theta$ between $40$ and $\unit[54]{^{\circ}}$
in $\unit[2]{^{\circ}}$ steps per pad (corresponding to $\unit[\approx
1.7]{mm}$).
\begin{figure}
  \includegraphics[width=\linewidth]{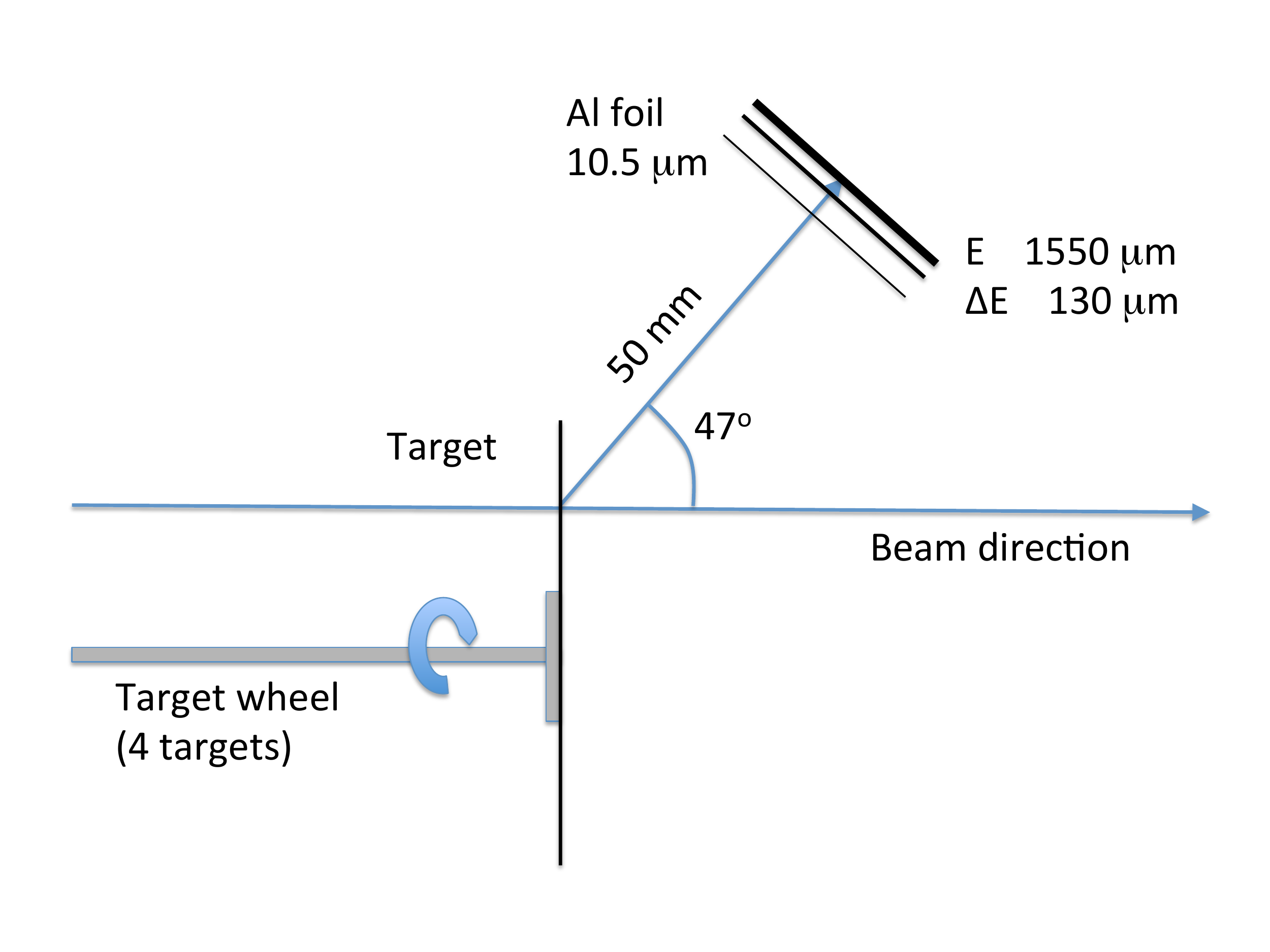}
  \caption{Illustration of the set-up. Only one $\Delta E-E$ detector
    module is shown with a center at $\theta = 47^{\circ}$ with
    respect to the beam axis. One cone of aluminum foil is placed in
    front of all the 8 telescope modules to reduce $\delta$-electrons
    impinging on the front detector.  The target chamber also houses a
    target wheel with place for 4 targets. }
  \label{fig:chamber}
\end{figure}
Figure~\ref{fig:chamber} shows the arrangement of the telescope system
within the target chamber.

The detector system is designed for measuring various outgoing charged
particles appearing for the projectile types and energies available at
OCL.
The yield of making good 2--\unit[4]{cm$^2$} area detectors with
thickness $\unit[> 2]{mm}$, is low due to bad bulk properties as a
result of an increasing number of impurities.
Also, high depletion voltages require that broad guard rings surround
the active areas.
A good compromise for the beam energies needed for the Oslo method, is
a $\Delta E$ and $E$ detector with thicknesses of $130$ and
$\unit[1550]{\upmu m}$, respectively.
Such a telescope system will be able to measure and identify protons
and \textsuperscript{4}He-ions in the energy regions of $3.7$ --
$\unit[16.5]{MeV}$ and $15.0$ -- $\unit[63.0]{MeV}$, respectively.
\begin{table}
  \caption{
    Particle energies deposited in the telescope.
    The second column gives the maximum energy deposited in the
    $\Delta E$ front detector, which represents the lowest
    energy applicable.
    The three columns to the right represent the highest energy
    that is stopped by the $\Delta E + E$ detector, and the
    corresponding energy deposits in the $\Delta E$ ($\unit[130]{\upmu m}$)
    and $E$ ($\unit[1550]{\upmu m}$) detectors.}
  \begin{tabular}{lcc c c}
    \hline
    \hline
    Particle   &$\Delta E$   &$\Delta E + E$&$\Delta E$ &    $E$  \\
    type       &  (MeV)      &    (MeV)     &    (MeV)  &   (MeV) \\
    \hline
    p          &     3.7     &      16.5    & 0.7      &   15.8 \\
    d          &     4.9     &      22.3    & 1.0      &   21.3 \\
    t          &     5.7     &      26.5    & 1.2      &   25.3 \\
    \textsuperscript{3}He
               &     13.4    &      58.3    & 2.6      &   55.7 \\
    $\upalpha$ &     15.0    &      65.9    & 2.9      &   63.0 \\
    \hline
  \end{tabular}
  \\
  \label{tab:tab1}
\end{table}
A more complete list of particle types and energies is shown in
Table~\ref{tab:tab1}.

\begin{figure}
  \includegraphics[width=\linewidth]{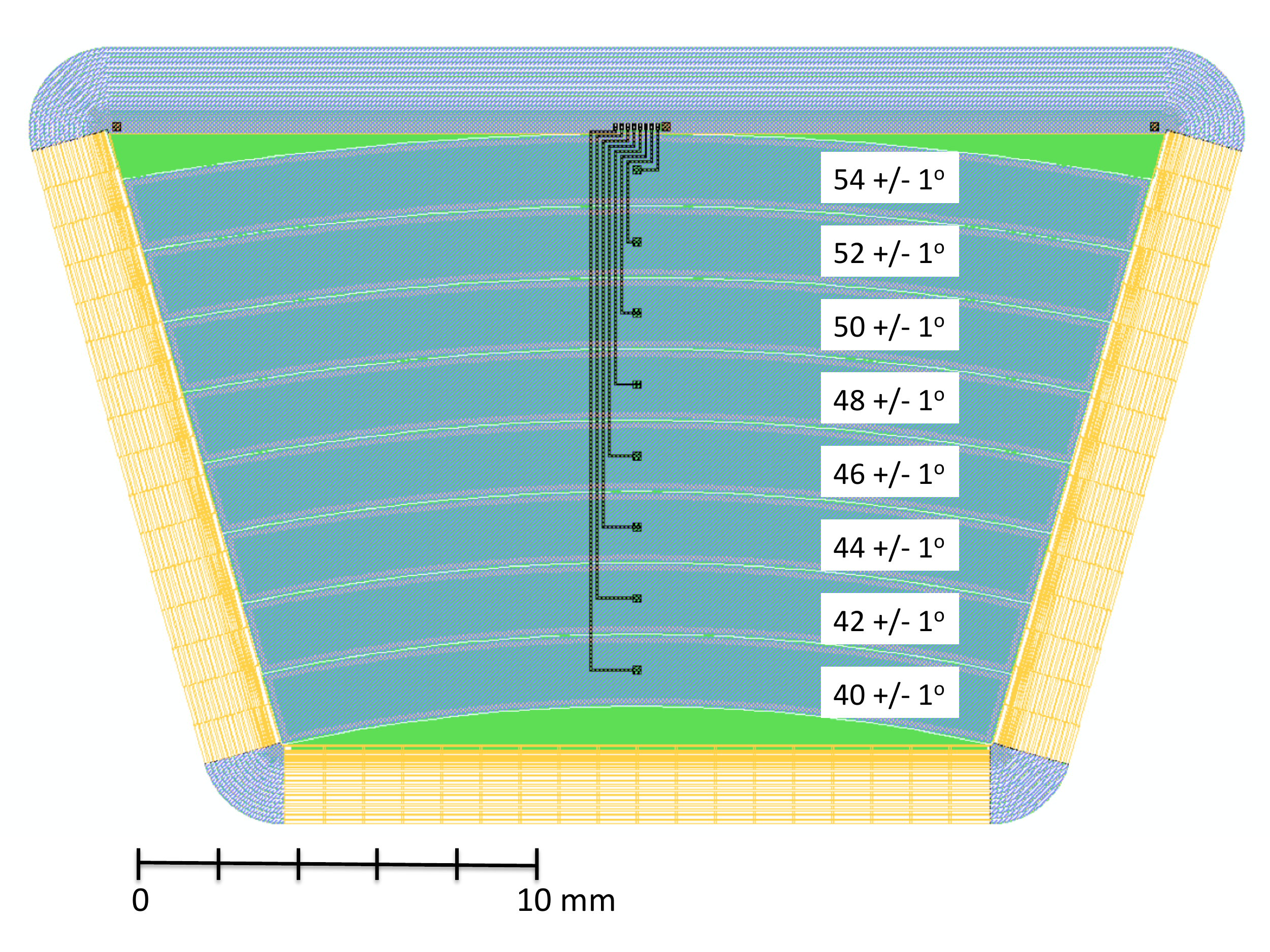}
  \caption{Layout of the front $\Delta E$ detector. The curved pads are
  designed to specific angles $\theta$. }
  \label{fig:trapeze}
\end{figure}

\section{Detector Layout}
\label{sec:layout}

The thick $E$ detector ($\unit[1550]{\upmu m}$) needs a high bias
voltage in order to be fully depleted.
Therefore, 18~guard rings are surrounding each detector's active area,
covering a ring width of $\unit[1700]{\upmu m}$, which is
comparable with the detector thickness.
As $\Delta E$ and $E$ detectors are mounted just behind each other, a
larger active area in the thin detectors would not increase the
efficiency of coincident $\Delta E-E$ measurements.
In order to avoid extra mask costs, it was therefore decided to equip
the $\Delta E$ detectors with the same guard-ring structure.

Figure~\ref{fig:trapeze} shows the layout of the thin $\Delta E$ front
detector.  The detector is equipped with eight curved pads so that the
scattering angle $\theta$ is constant for each pad.
Due to this curvature and the trapezoidal shape of the detector, an
area about as large as half a pad is not used for detection, see
Fig.~\ref{fig:trapeze}.
The area of the pads increases with $\theta$.
In the spherical limit (ignoring the guard rings), the corresponding
solid angle covered by each pad is
\begin{equation} 
  \Delta \Omega = 2 \pi \sin \theta \Delta \theta.
\end{equation}
Thus, the solid angle covered by the $40^{\circ}$ pad is about
$\unit[21]{\%}$ smaller than for the $54^{\circ}$ pad.
The back $E$ detector has the same layout as shown in
Fig.~\ref{fig:trapeze}, but is not segmented into several pads.

The $\Delta E$ and $E$ detector chips were designed and produced by
SINTEF MiNaLab, Norway.
Float zone (FZ) silicon originating from Topsil, Denmark, was used in
the production.
The wafers for the $\unit[1550]{\upmu m}$ thick $E$ detector were
supplied directly by Topsil, while the $\unit[130]{\upmu m}$ thick
wafers for the $\Delta E$ detector were procured from Virginia
Semiconductor, USA, who made the wafers from a FZ Topsil ingot.

\begin{figure}
  \includegraphics[width=\linewidth]{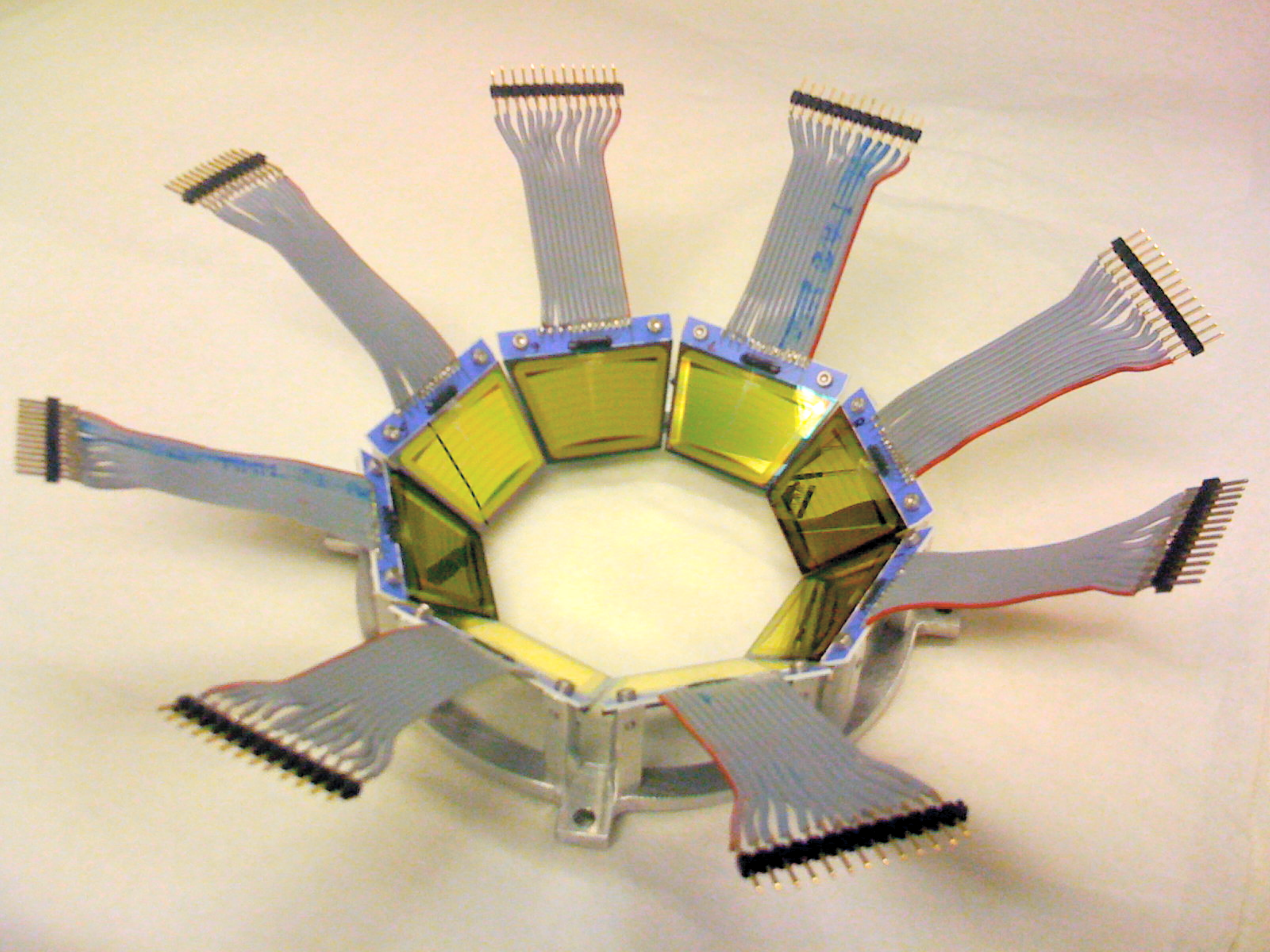}
  \caption{Silicon particle telescope modules with connectors,
    mounted on the support structure centering the detectors in the
    reaction chamber.}
  \label{fig:module}
\end{figure}

The processing sequence includes field oxidation, boron implantation
for the detector readout pads and guard ring, opening of contact
holes, and front and backside metalization (aluminum).
As the detector readout pads are covered by aluminum, the design of
the $\Delta E$ chip with eight pads requires a second layer of
aluminum.
This is necessary for crossing the lines connecting to the respective
bonding pads over the other readout pads.
The two metal layers are separated and isolated by $\unit[5]{\upmu m}$
of polyimide, and five mask layers are therefore needed for the
processing (active pad and guard ring, contact holes, metal~1,
polyimide, and metal~2).
As the $E$ detector chip only includes one readout pad, no second
metal is needed, and the processing requires three mask layers only.

The detector full depletion voltage is inversely proportional to the
specific resistivity, but increases with the square of the thickness.
The thick wafers used for production of the $E$ detector had a
specific resistivity in the range $10$ -- $\unit[30]{k\Omega cm}$.
The detectors are to be operated fully depleted, and the typical
depletion voltage was measured to $\unit[< 300]{V}$.
Another challenge is that the bulk leakage current increases with the
depletion width and thereby the thickness.
However, SINTEF has developed very efficient gettering processes which
eliminates most of the bulk recombination centers, and typical pad and
guard ring leakage currents at $\unit[480]{V}$ were $\unit[< 5]{nA}$
and $\unit[<10]{nA}$, respectively.
Concerning the $\Delta E$ detector, the main problem was the fragility
with resulting wafer breakage due to the very thin material and
especially insufficient edge rounding.

\begin{table}
  \caption{Silicon chip properties.}
  \begin{tabular}{lcc}
    \hline
    \hline
    Detector type            &   $\Delta E$            &       $E$    \\
    Chip \#                  &       21                &      23/5    \\
    \hline
    Thickness (mm)           &     0.13                &       1.55   \\
    Number of pads           &      8                  &        1     \\
    Pad area (mm$^2$)        &     299                 &       323    \\
    Individual pads (mm$^2$) &  31.5 - 43.7            &         -    \\
    Depletion (V)            &      15                 &       220    \\
    Pad leakage (nA)         &     0.4 @ 30V           &  6.5 @ 480V  \\
    Guard leakage (nA)       &     0.9 @ 30V           &  7.3 @ 480V  \\
    \hline
  \end{tabular}
  \label{tab:tab2}
\end{table}
Table~\ref{tab:tab2} shows typical depletion voltages and leakage
currents for the detectors.

The bonding and mounting on ceramic substrate were performed by
Microcomponent, Horten.
The two $\Delta E$ and $E$ chips are glued back-to-back on the
\unit[0.5]{mm} thick substrate.
To assure redundancy, two bonding threads were used for each contact
to the ceramic board.
A flat cable is soldered to the board to connect to the preamplifiers.
The assembled SiRi $\Delta E-E$ ring with~8 modules is shown in
Fig.~\ref{fig:module}.

\section{Electronics and Data Acquisition}
\label{sec:electronics}

The telescope module of Fig.~\ref{fig:module} is connected by
multi-pole shielded cables, manufactured by Mesytec, with LEMO vacuum
feedthroughs.
Outside the vacuum chamber, the detectors signals are connected to
preamplifiers.
There are four preamplifiers for the $\Delta E$ detectors, each
handling~16 pads, and one preamplifier for all eight $E$ detectors.
Both preamplifier types are Mesytec MPR-16 with sensitivities adapted
to the expected energy deposits in the front and back detectors,
respectively.

The preamplified signals are transferred as differential signals to
Mesytec STM-16 modules including both spectroscopy amplifiers and
timing-filter amplifiers, and also leading-edge discriminators.
The logic or of all $E$ detector discriminator outputs is used to
generate the trigger signal for the data acquisition.

\begin{figure}
  \includegraphics[width=\columnwidth]{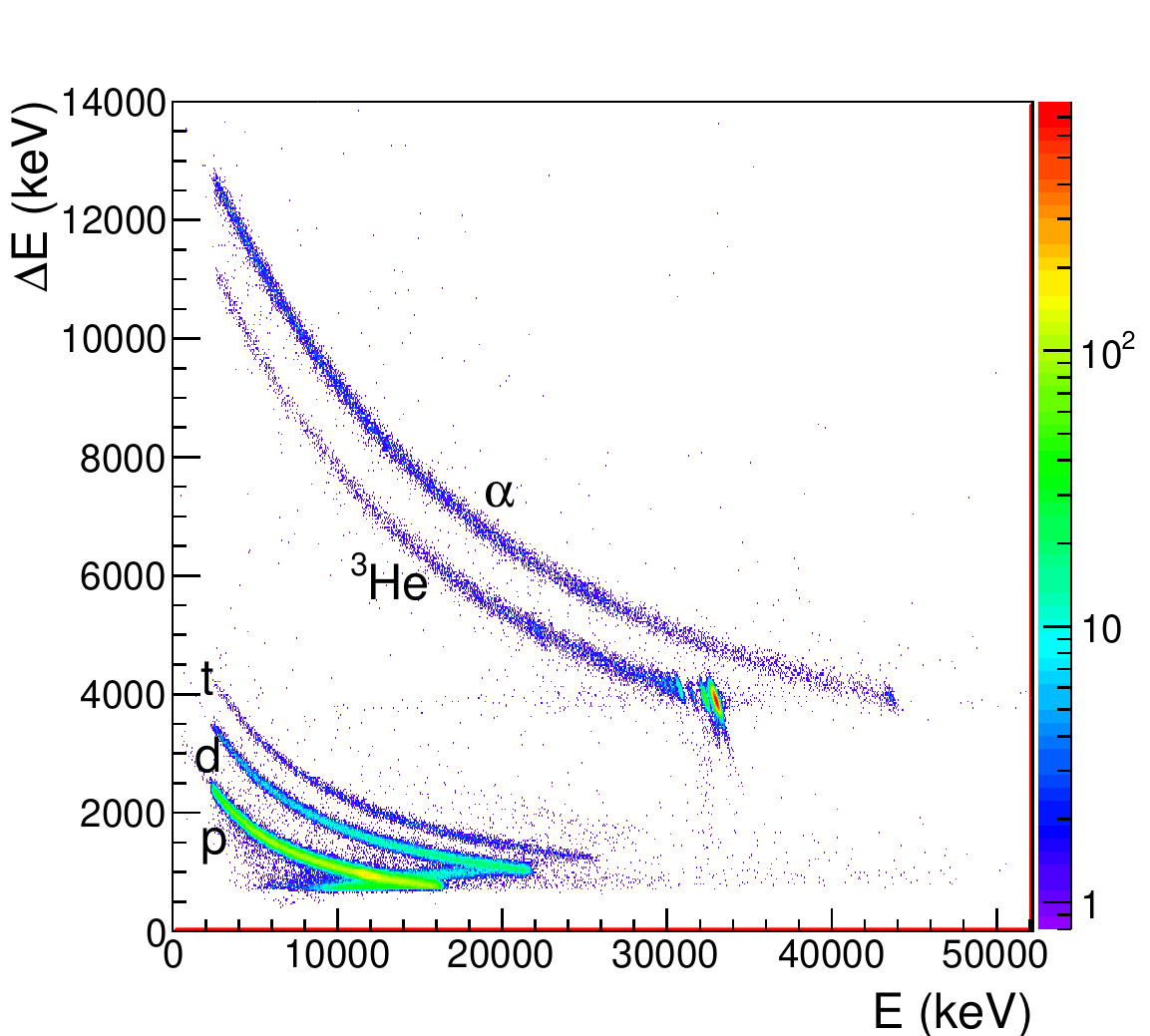}
  \caption{$\Delta E - E$ matrix for the reaction of $\unit[38]{MeV}$
    \textsuperscript{3}He ions on a \textsuperscript{112}Cd
    target. For this example, we have chosen front detector f5
    ($\theta = 50^{\circ}$) and back detector b1.  There are totally
    64 matrices with f0, f1, \dots, f7 and b0, b1, \dots, b7.}
  \label{fig:banana}
\end{figure}

The $\gamma$-rays detected by CACTUS are filtered off-line to select
only those rays in coincidence with the respective reaction of
interest.
This is achieved by measuring the time difference between particle
detection in the $E$ detector (start signal) and the $\gamma$-ray
detection in CACTUS (stop signal).
The acquisition trigger signal is given by the logic OR of all $E$
detector discriminator outputs, optionally AND-ed with the logic OR of
all $\Delta E$ detector discriminator outputs.
The stop signal is individual for each $\gamma$-ray detector, i.e.,
for 28~NaI and up to 2~Ge detectors.

Since we use leading-edge and not constant-fraction discriminators,
the walk due to different signal rise times for different energy
deposits has to be corrected in software.
For this purpose, we found that a good choice for the energy-corrected
time was given by
\begin{equation}
  t(E) = t_0 + \frac{\alpha}{E+\beta}+\gamma E,
  \label{eq:time}
\end{equation}
where $t_0$ is the measured time and $\alpha$, $\beta$ and $\gamma$
are fitted values to ensure that $t(E)$ is approximately constant.

The data acquisition system is based on one VME crate housing
commercial and custom-made VME modules.
The system is controlled by software running on a CES 8062 CPU.
The trigger handling is performed by a custom VME module which is
capable of separating~8 different trigger sources.
The analog-to-digital conversion is done using ADCs from CAEN
(mod.~785) and Mesytec (MADC-32), and TDCs from CAEN (mod.~775).
The data is transferred to a standard Linux PC through a CAEN VME USB
module.
The whole system has been run without problems at trigger rates of up
to $\unit[10]{kHz}$.

The slow-control settings of most Mesytec modules are operated via
Mesytec's proprietary remote control bus using a control software
developed at OCL.
This remote control is very convinient for modules placed at the
target station (ramping of HV and leakage current monitoring, no
radiation exposure), as well as for the shaper modules (thresholds and
gains, large number of channels to adjust).
The thresholds and control registers of the ADCs and TDCs are set
directly by the data aquisition program running on the VME CPU.

\section{System performance}
\label{sec:performance}

The new SiRi particle-telescope system has already been used in
several experiments at OCL.
In principle there is no need for constructing a fast coincidence
overlap between the $\Delta E$ and $E$ detectors.
If one back $E$-trapeze has triggered, also the front detector should
have been hit by the same charged particle, unless the particle passed
through the areas not covered by the strips.
By requiring that one and only one pad of the front detector has
provided a reasonably high signals, the $\Delta E-E$ particle event is
assumed to be good.

Figure~\ref{fig:banana} shows a typical $\Delta E-E$ matrix for
$\unit[38]{MeV}$ \textsuperscript{3}He ions impinging on a
\textsuperscript{112}Cd target.
The curves for each particle type are well separated, and the
coincident $\gamma$ rays can be assigned to a specific nucleus at a
given excitation energy $E_x$, with $E_x <B_n$.
The most energetic protons, deuterons, and tritons are not stopped in
the $E$ detector, resulting in a backbend of the respective curves.

\begin{figure}[thb]
  \includegraphics[width=\columnwidth]{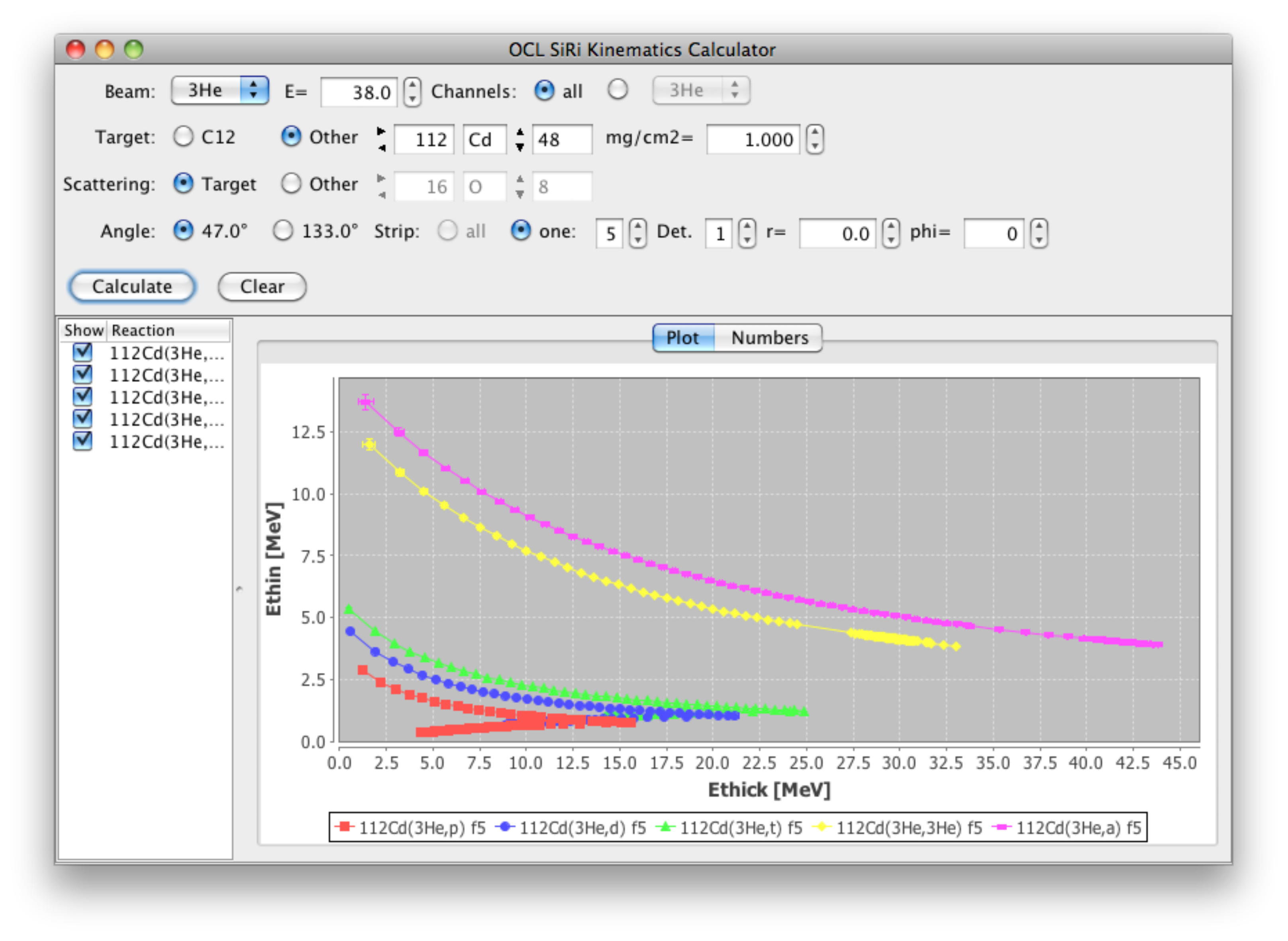}
  \caption{Graphical user interface (GUI) of the jkinz application
    with parameters appropriate for the \textsuperscript{112}Cd
    experiment of Fig.~\ref{fig:banana}.}
  \label{fig:banana_theo}
\end{figure}

A computer code jkinz~\cite{jkinz} has been developed to calculate
reaction kinematics and to estimate the energy losses of the various
particle types in the target and other materials.
The energy loss functions by Ziegler~\cite{Ziegler} are used for this
purpose.
The nuclear masses necessary for the relativistic treatment of the
reaction kinematics are obtained from the AME2003
tables~\cite{AME2003}.
The calculation displayed in Fig.~\ref{fig:banana_theo} demonstrates
the very good resemblance with the experimental curves of
Fig.~\ref{fig:banana}.

\begin{figure}
  \includegraphics[width=0.7\linewidth]{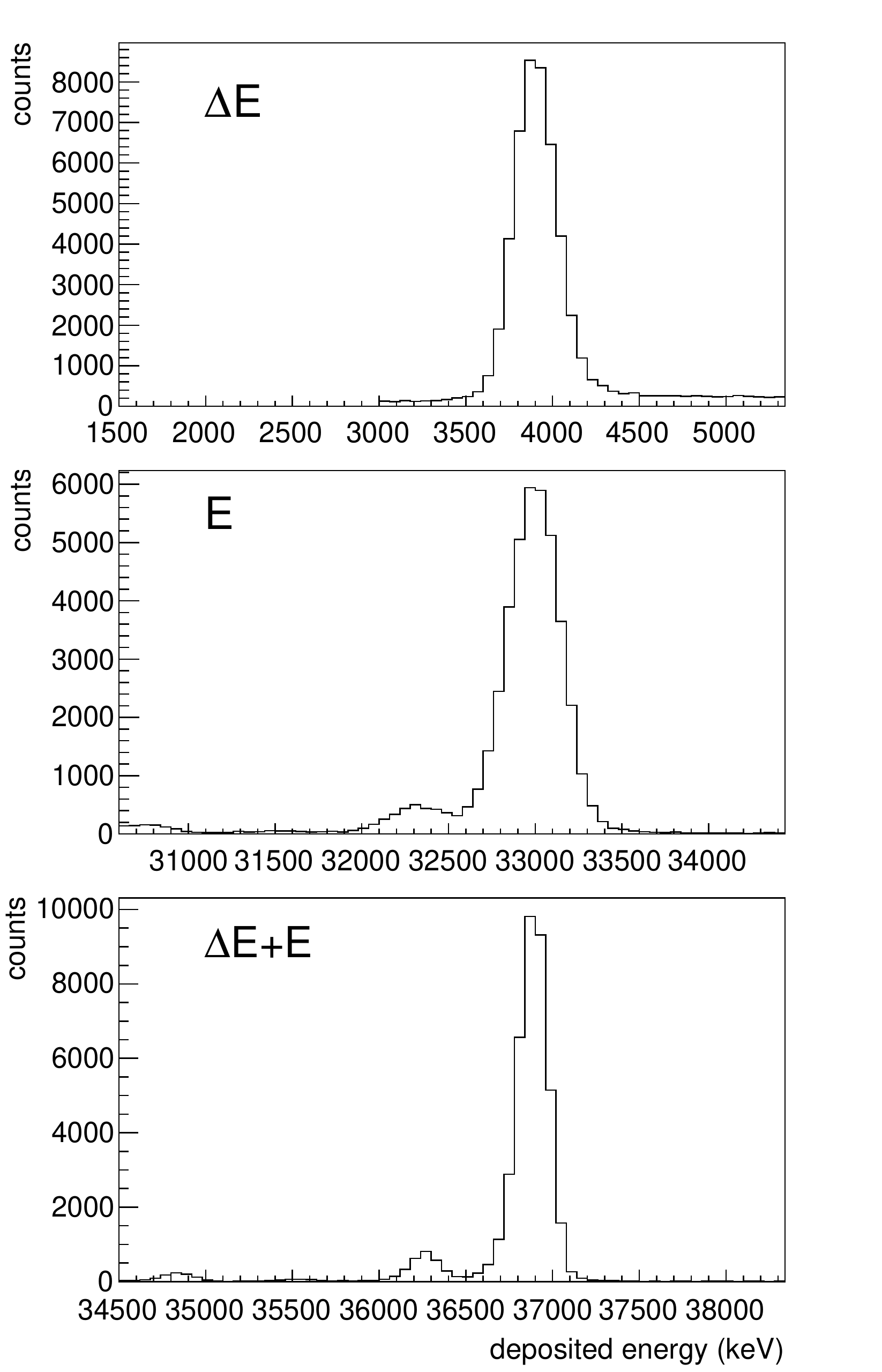}
  \caption{Spectra of the measured
    \textsuperscript{112}Cd(\textsuperscript{3}He,\textsuperscript{3}He)\textsuperscript{112}Cd
    elastic peak in the $\Delta E$ and $E$ detector. The bin width is
    60 keV/ch. A clear improvement in energy resolution is seen in the
    spectrum where $\Delta E$ and $E$ are added event-by-event.}
  \label{fig:elastic}
\end{figure}

\begin{figure}
  \includegraphics[width=\linewidth]{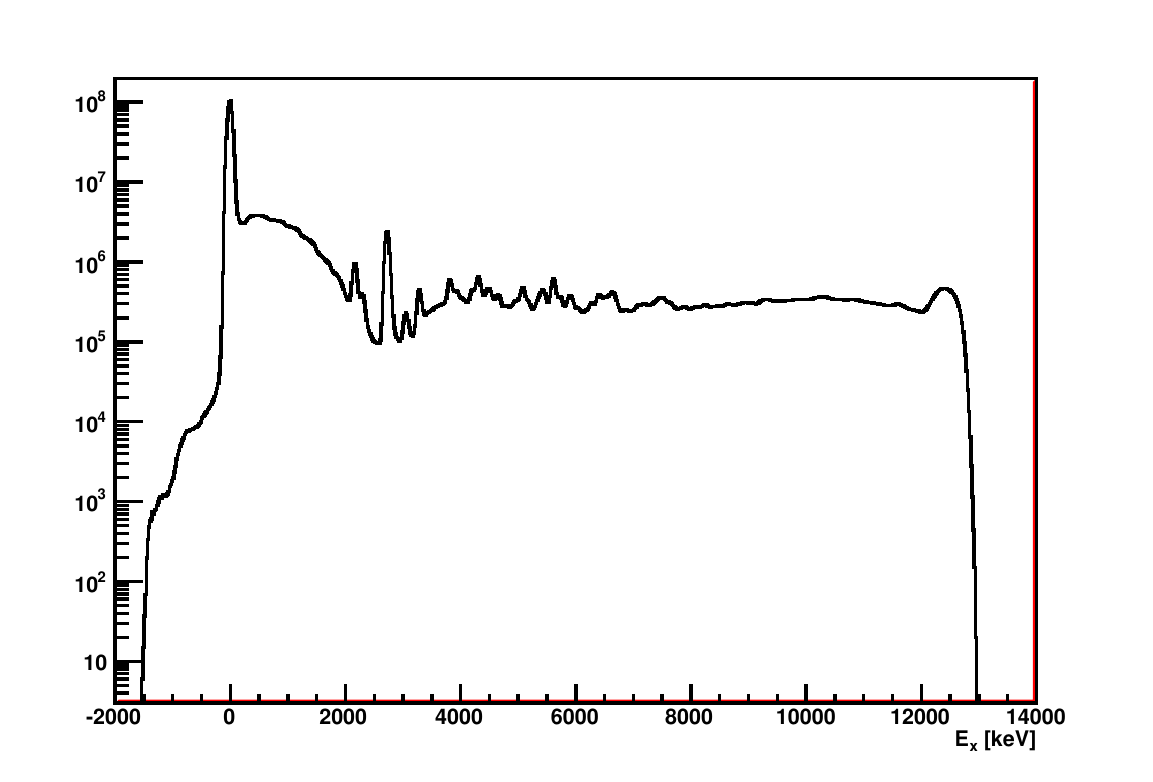}
  \caption{Proton spectrum of the
    \textsuperscript{90}Zr$(p,p')$\textsuperscript{90}Zr reaction with
    beam energy of $\unit[17]{MeV}$. All 64 particle telescopes are added.}
  \label{fig:90zr}
\end{figure}

Projections of the \textsuperscript{3}He curve of
Fig.~\ref{fig:banana} on the $\Delta E$ and $E$ axis are shown in
Fig.~\ref{fig:elastic}.
The spectra are displayed for energies around the elastic peak.
The spectrum created event-by-event by adding the two detector signals
$E_\text{tot}= \Delta E + E$ gives a resolution which is about two
times better than for the $E$ projection.
The reason is that the more energy deposited in the $\Delta E$
detector, due to statistical straggling, the less energy is deposited
in the $E$ detector, and opposite.
The FWHM of the elastic scattering peak in the $E_\text{tot}$ spectrum
is approximately $\unit[200]{keV}$, which is very good with respect to
all contributing factors.
The excited $2^{+}$ state of \textsuperscript{112}Cd at
$\unit[618]{keV}$ is well separated from the strong elastic peak.

The main contribution to the total resolution of the $E_\text{tot}$
spectra has its origin from the variation of recoil energy carried by
the heavy residual nucleus; the higher scattering angle $\theta$, the
more kinetic energy is transferred to the residual nucleus.
This effect can be reduced by lighter projectile with lower incident
energy, and/or by using heavier targets.

\begin{figure}[thb]
  \includegraphics[width=\linewidth]{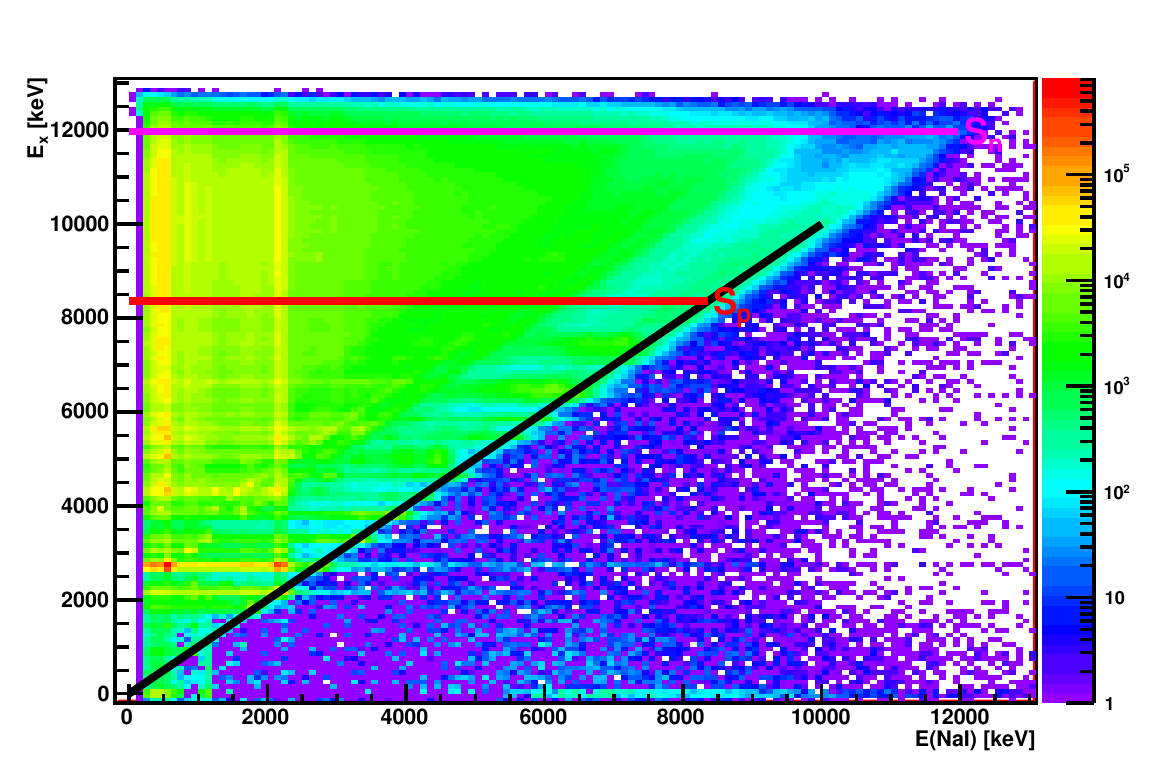}
  \caption{Proton-$\gamma$ coincidences giving the $E_x-E_{\gamma}$ matrix,
  which is the starting point for the Oslo method. It should be noted that 
  the NaI spectra are raw, meaning they have not been unfolded 
  by the NaI detector responce function. The horizontal lines marked $S_p$ and $S_n$ indicate the proton and neutron binding energies, respectively.}
  \label{fig:alfna}
\end{figure}

Figure~\ref{fig:90zr} shows the results from a typical light-ion
experiment~\cite{Zr9092Paper} with $\unit[17]{MeV}$ protons on
\textsuperscript{90}Zr.
The experimental resolution for the ground state in $(p,p')$
scattering on $\unit[1.83]{mg/cm^2}$ \textsuperscript{90}Zr is now
FWHM~$\unit[\approx100]{keV}$, corresponding to a standard deviation
of $\sigma\approx\unit[43]{keV}$.
This resolution includes the straggling in the target and the
uncertainty in the scattering angle determination.
It also includes all misalignments of the detector system.

The elastic peak is seen to be more than 100~times stronger than the
average (p,p$^\prime$) cross-section to excited states in
\textsuperscript{90}Zr.
The rate of pile-up events is 4~orders of magnitude lower than the
elastic peak.
The particle yield at the right-hand tail of the elastic peak is due
to $\approx 20$\% punch-through of the elastic events.

A good SiRi particle event is to be taken in coincidence with the NaI
and Ge detectors of the CACTUS array.
Here, the 32-fold TDC gives the time difference between the $E$
detector and the individual $\gamma$ detectors.
In the event sorting procedure, the energy-compensated time difference
is reconstructed by
\begin{equation}
  \Delta t(E_{\rm back},E_{\gamma}) = \Delta t_0 -t_{\rm p}(E_{\rm back})-t_{\gamma}(E_{\gamma}),
\end{equation}
where the two last terms are calculated from Eq.~(\ref{eq:time}).
The two sets of $\alpha$, $\beta$ and $\gamma$ parameters needed, were
fitted to data from a separate run on a \textsuperscript{12}C target.
In practice, it is usually sufficient to set each NaI detector's $t_0$
value such that all detectors are aligned at $E_{\gamma}=4.43$~MeV,
and then use the same energy-dependent correction to all NaI
detectors, as the output signal amplitudes of the NaI detectors are
usually adjusted to be very similar to each other.
A similar procedure is applied for the time signals of the $E$
detectors.
Here, the corrections are usually small because of the fast signals
from the $E$ detector and the high ejectile energy and small energy
interval ($E_x < 8$ -- $12$ MeV) of interest.

For low energy signals, $\alpha$ is the most important parameter
describing the hyperbolic energy dependence of the trigger time close
to the energy threshold.
Here, we find $\alpha < 0$ for the START $E_{\rm back}$ detector and
$\alpha > 0$ for the STOP $\gamma$ detectors since the low energy
signals produce delayed leading-edge discriminator triggers.
The procedure for making energy-compensated time spectra works very
good and the resulting total time resolution of 8~particle telescopes
and 28~NaI detectors is about FWHM~$=15$ -- $\unit[20]{ns}$.
The main contribution to the resolution comes from the NaI PMTs, which
are optimized for good energy resolution, and not time.

Figure~\ref{fig:alfna} shows the results from the particle-$\gamma$
coincidence measurement.
The relation between particle energy and excitation energy is
established using calculations performed with the jkinz application,
so that the excitation energy can be deduced from the particle energy.
A prompt time gate is set on the coincidence peak of the $\Delta
t(E_{\rm back},E_{\gamma})$ spectrum for incrementing the
($E_{\gamma},E_x$) entries event-by-event, and a time gate on the
random coincidences is set for decrementation.
Also a gate on the proton particle $\Delta E-E$ curve is required to
reduce the occurrence of unwanted events originating from pile-up,
$\delta$-electrons, incomplete energy deposits, channeling effects in
silicon and so on.

The data of Fig.~\ref{fig:alfna} fall mostly within the triangle
defined by $E_{\gamma}<E_x$.
The small number of counts outside this triangle shows that the
coincidences are true and the pile-up is small.
Some $\gamma$-ray lines are seen as vertical lines.
They represent yrast transitions passed in almost all cascades for a
large range of initial excitation energies, up to the neutron
separation energy of $E_x=S_n\approx \unit[12]{MeV}$.

\section{Conclusion}
\label{sec:conclusion}

The SiRi particle-telescope system has been used in various
experiments at the Oslo Cyclotron Laboratory.
The system is able to identify the charged particle type using the
well-known $\Delta E-E$ curve gating technique.
The particle resolution is better and the efficiency is about 10~times
higher than with the previous set-up of conventional silicon
detectors.

SiRi also allows to study ejectiles in 8~angles with $\theta=40$ --
$54^{\circ}$ relative to the beam direction, and 8~angles around the
beam axis with $\phi = 0$ -- $360^{\circ}$.
This gives the opportunity to explore the angular momentum transfer in
the direct reactions.

The combined SiRi-CACTUS system has also been tested, giving very nice
particle$-\gamma$ coincidences.
The random coincidences are subtracted in a satisfacory way, and the
measurements are not affected by severe pile-up effects, provided that
the beam current is typically less than $\approx \unit[2]{nA}$.
By utilizing the ejectile-$\gamma$-ray angular coorelations,
information on the multipolarities of the $\gamma$ transitions should
be possible to deduce as function of initial excitation energy.

We believe that the good-resolution, high-efficiency particle-$\gamma$
coincidence system will open for the study of new physics in the
quasi-continuum of atomic nuclei.

\acknowledgments

Financial supports from the Norwegian Research Council (NFR) and the
University of Oslo are gratefully acknowledged.
We also thank A.~Schiller and A.~Werner for their contribution in the
early stage of the project, and A.C.~Larsen for preparation of
Figs.~\ref{fig:banana} and \ref{fig:elastic}.


\begin{thebibliography}{99}

\bibitem{OsloMethod}A.~Schiller, L.~Bergholt, M.~Guttormsen, E.~Melby,
  J.~Rekstad, and S.~Siem, Nucl.\ Instrum.\ Methods Phys.\ Res.\ A \bf
  447\rm, 498 (2000).

\bibitem{SnPaper} U.~Agvaanluvsan, A.C.~Larsen, M.~Guttormsen,
  R.~Chankova, G.~Mitchell, A.~Schiller, S.~Siem, and A.~Voinov,
  Phys.\ Rev.\ C \bf 79\rm, 014320 (2009).

\bibitem{Sc45Paper} A.C.~Larsen, M.~Guttormsen, R.~Chankova,
  T.~L\"onnroth, S.~Messelt, F.~Ingebretsen, J.~Rekstad, A.~Schiller,
  S.~Siem, N.U.H.~Syed, and A.~Voinov, Phys.\ Rev.\ C \bf 76\rm,
  044303 (2007).

\bibitem{Sc43Paper} A.~B\"urger, S.~Hilaire, A.C.~Larsen, N.U.H.~Syed,
  M.~Guttormsen, S.~Harissopulos, M.~Kmiecik, T.~Konstantinopoulos,
  M.~Krti\v{c}ka, A.~Lagoyannis, T.~L\"onnroth, K.~Mazurek, M.~Norby,
  H.~Nyhus, G.~Perdikakis, S.~Siem, and A.~Spyrou, Phys.\ Rev.\ C, in
  preparation
  
\bibitem{jkinz} A.~B\"urger, OCL SiRi Kinematics Calculator, University of Oslo, 2011,
  \url{http://tid.uio.no/~abuerger/}

\bibitem{Ziegler} J.F.~Ziegler, J.P.~Biersack, and U.~Littmark,
  \textit{The Stopping and Range of Ions in Solids}, Pergamon Press,
  New York (1985)

\bibitem{Zr9092Paper} A.~B\"urger, S.~Siem, A.~G\"orgen,
  M.~Guttormsen, T.W.~Hagen, A.C.~Larsen, P.~Mansouri, M.H.~Miah,
  H.T.~Nyhus, Th.~Renstr{\o}m, S.J.~Rose, N.U.H.~Syed, H.K.~Toft,
  G.M.~Tveten, A.~Voinov, and K.~Wikan, Phys.\ Rev.\ C, in preparation

\bibitem{Sn121122Paper} H.K.~Toft, A.C.~Larsen, A.~B\"urger, {\it et
    al.}, PRC, accepted (2011)

\bibitem{AME2003} G.Audi, A.H.Wapstra, and C.Thibault, Nucl.\ Phys\
  A\textbf{729}, 337--676 (2003)

\end{thebibliography}
\end{document}